\documentclass[prb,
twocolumn,
superscriptaddress,showpacs,amsmath,amssymb]{revtex4}
\usepackage{graphicx}

\begin{document}

\title{Type-II Weyl semimetal vs gravastar}

\date{\today}

\begin{abstract}
The boundary between the type I and type II Weyl semimetals serves as the event horizon for the "relativistic" fermions.
 The interior of the black hole is represented by the type II Weyl semimetal, where the Fermi surface is formed.
 The process of the filling of the Fermi surface by electrons results in the relaxation inside the horizon. This leads to the Hawking radiation and to the reconstruction of the interior vacuum state. After the Fermi surface is fully occupied, the interior region reaches the equilibrium state, for which the Hawking radiation is absent.
If this scenario is applicable to the real black hole, then the final state of the black hole will be the dark energy star with the event horizon.
Inside the event horizon one would have de Sitter space time, which is separated from the event horizon by the shell of the Planck length width. 
Both the de Sitter part and the shell are made of the vacuum fields without matter.  This is distinct from the gravastar, in which the matter shell is outside the "horizon", and which we call the type I gravastar.
But this is similar to the other type of the vacuum black hole, where the shell is inside the event horizon, and which we call the type II gravastar.
We suggest to study the vacuum structure of  the type II gravastar using the $q$-theory, where the vacuum variable is the 4-form field introduced for the phenomenological description of the quantum vacuum.
 \end{abstract}
\pacs{
}

\author{G.E.~Volovik}
\affiliation{Low Temperature Laboratory, Aalto University,  P.O. Box 15100, FI-00076 Aalto, Finland}
\affiliation{Landau Institute for Theoretical Physics, acad. Semyonov av., 1a, 142432,
Chernogolovka, Russia}

\maketitle

% \tableofcontents
 \newpage

%\twocolumn

\section{Introduction}

 There are different scenarios of the development of the black hole in the process of evaporation by Hawking radiation. In particular, the end of the evaporation can result in a macroscopic quantum tunnelling from the black hole to the white hole, see Ref.\cite{Rovelli2021} and references therein.
 Another scenario is the formation of compact object -- the vacuum star.
In particular, the black hole event horizon can be considered as the boundary separating different phases of the quantum vacuum.\cite{Chapline2003}  Such consideration was based on the condensed matter analogies, which in particular 
are presented by the superfluid phases of liquid $^3$He.\cite{Volovik2001}
Topological materials such as the Weyl and Dirac semimetals bring a new twist to this analogy. 
They provide new support for the scenario of the formation of a vacuum star after the end of Hawking evaporation.

The analog of the event horizon emerging on the boundary between type I and type II Weyl semimetals is considered in Section \ref{WeylSection}.
In Section \ref{VacuumStarSec} we compare the black hole in semimetals with different types of the vacuum stars (gravastars).
The comparison gives the preference to the gravastars with well defined event horizon. which we call the type II gravastar.
This compact object contains three regions with different realizations of the quantum vacuum. These are: the Sitter vacuum inside the  Cauchy horizon, the vacuum in the thin shell separating the Cauchy horizon from the event horizon, and the vacuum outside of the event horizon.
In Section \ref{VacuumQSec} there is an attempt to describe  all three regions using the phenomenological theory of the quantum vacuum in terms of the 4-form field.\cite{Duff1980,Aurilia1980,Hawking1984}

\section{Analog of horizon in Weyl semimetals}
\label{WeylSection}

\begin{figure}
\centerline{\includegraphics[width=\linewidth]{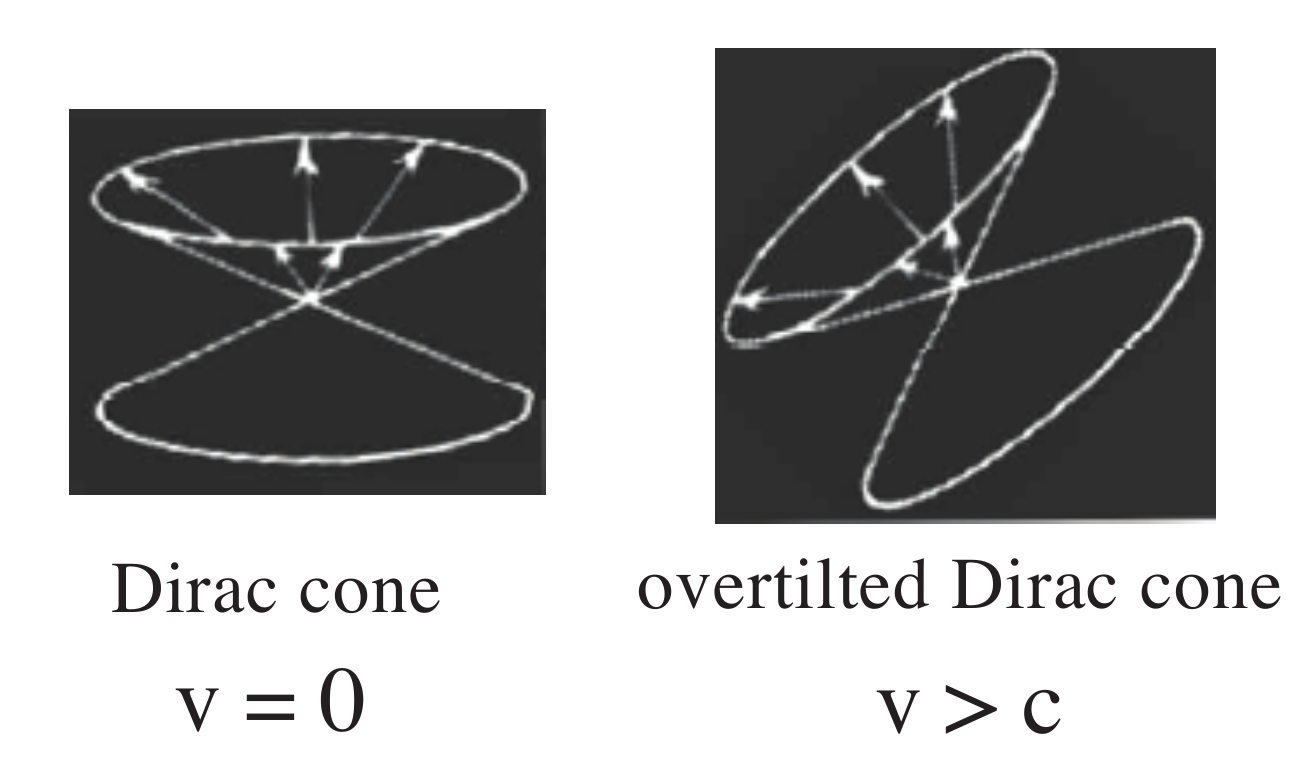}}
\caption{ The energy spectrum  in Weyl semimetals, which follows from Eq.(\ref{EffectiveMetricC}), is $E({\bf p}) = p_r v \pm pc$.
It forms the Dirac cone.  \\
{\it Left}: In  the type I Weyl semimetal one has $|v|<1$. Figure shows the Dirac cone with $v=0$. For $0<|v| <1$ the cone is tilted. 
\\
{\it Right}: For  $|v|>1$ the Dirac cone is overtilted. This is called the type II Weyl semimetal. The overtilted Dirac cone crosses the zero energy level, which gives rise to the Fermi surface $E({\bf p})=0$, which is conical, $p_r/p= c/v<1$. At large momenta, far away from the Weyl point, the Fermi surface becomes closed due non-linear corrections to the linear spectrum, see Fig. \ref{fig:FS}.  
}
\label{fig:cone}
\end{figure}

The energy spectrum of electrons in Weyl semimetals becomes "relativistic" in the vicinity of the Weyl point. In its simplest form the Hamiltonian near the Weyl point at ${\bf p}=0$ is:
 \begin{equation}
H({\bf p}, {\bf r})= c \boldsymbol\sigma \cdot {\bf p} + {\bf v}({\bf r}) \cdot {\bf p}\,,
\label{WeylHamiltonian}
\end{equation}
where $\boldsymbol\sigma$ are the Pauli matrices; $c=1$ is the analog of the speed of light.
In the more general form, the Weyl equation $-i\partial_t \Psi = H\Psi$ for the Weyl spinor $\Psi$ is written in terms of tetrads $e_a^\mu$, as the parameters of expansion at the Weyl point:
 \begin{equation}
e_a^\mu  \sigma^a p_\mu \Psi=0\,\,,  \,\, \sigma^a=(\boldsymbol\sigma,1) \,\,, \, \,p_\mu = -i \partial_\mu \,\,\,.
\label{WeylTetrad}
\end{equation}
The tetrad fields describe the effective gravity experienced by the Weyl fermions. The effective metric in this tetrad gravity is the secondary object obtained as the bilinear combination of the original tetrads, $g^{\mu\nu}=e_a^\mu e_b^\nu \eta^{ab}$. In the simplest case of Eq.(\ref{WeylHamiltonian}),
 the vector ${\bf v}({\bf r})$ is the analog of the shift velocity in this effective metric:\cite{Volovik2016}
 \begin{eqnarray}
 g^{\mu\nu} p_\mu p_\nu=-(p_0 -  {\bf v}({\bf r}) \cdot {\bf p})^2 + {\bf p}^2 =0 \,\,,\, p_0=E\,,
 \label{EffectiveMetricC}
 \\
ds^2=g_{\mu\nu} dx^\mu dx^\nu= - dt^2 +   (d{\bf r} - {\bf v}({\bf r})dt)^2\,.
\label{EffectiveMetric}
\end{eqnarray}
Here the contravariant metric $g^{\mu\nu}$ describes the Dirac cone in momentum space in Fig. \ref{fig:cone}, while the covariant metric $g_{\mu\nu}$ describes the corresponding light cone in the effective spacetime.
This metric is similar to the acoustic metric,\cite{Unruh1981} which emerges in moving liquids and superfluids, where the velocity of fluid plays the role of the shift velocity. But  the analog of shift velocity in Weyl semimetals is not related to any real motion of the substance, and this leads to the important consequences.

%%%%%%%%%%%%%%%%%%%%%%%%%%%%%%%%%%%%%%%%%%%%%%%%%%%%%%%%%
%%%%%%%%%%%%%%%%%%%%%%%%%%%%%%%%%%%%%%%%%%%%%%%%%%%%%%%%%

In the Weyl semimetals, the black hole horizon and the corresponding Hawking radiation can be simulated by creation of the region with overtilted Dirac cone,\cite{Volovik2016,Zhang2017,Ojanen2019,Wilczek2020,Hashimoto2020,DeBeule2021,Sabsovich2021,Morice2021} where $|{\bf v}({\bf r})| >1$ (see Fig. \ref{fig:cone} {\it right}). The event horizon is situated at the boundary between the region with $|{\bf v}({\bf r})| <1$, which is called the type I Weyl semimetal, and the region with the overtilted Dirac cone, $|{\bf v}({\bf r})| >1$, which is called the type II Weyl semimetal,\cite{VolovikZubkov2014} see also Ref. \cite{Hasan2021}.
For the spherical horizon, the corresponding effective metric has the Painlev\'e-Gullstrand (PG) form:\cite{Painleve,Gullstrand}
 \begin{equation}
ds^2= - dt^2 +   (dr - v(r) dt)^2+r^2 d\Omega^2\,.
\label{PG}
\end{equation}

%%%%%%%%%%%%%%%%%%%%%%%%%%%%%%%%%%%%%%%%%%%%%%%%%%%%%%%%%
%%%%%%%%%%%%%%%%%%%%%%%%%%%%%%%%%%%%%%%%%%%%%%%%%%%%%%%%%
\begin{figure}
\centerline{\includegraphics[width=\linewidth]{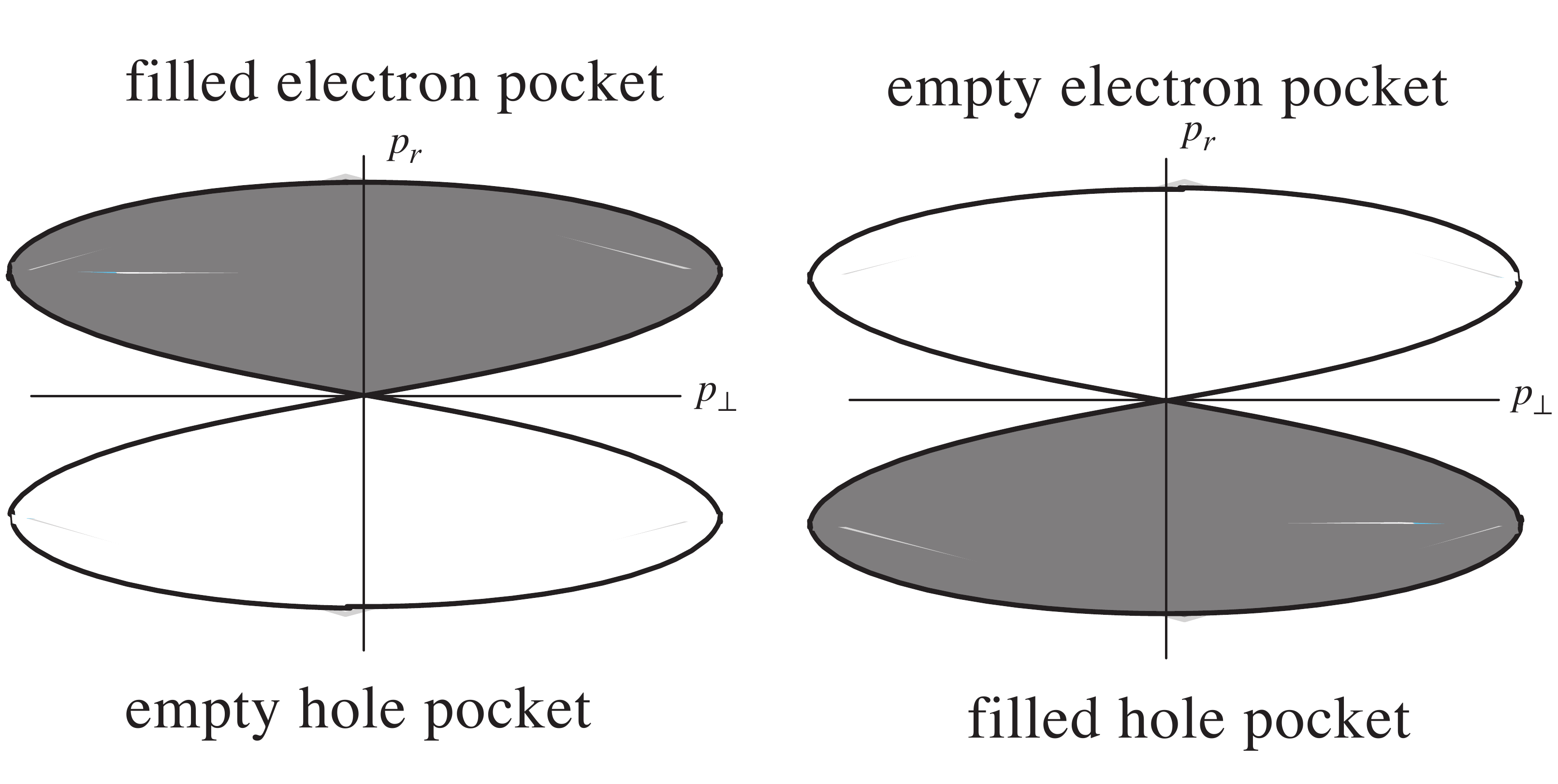}}
\caption{Fermi surface $E({\bf p})=0$ in the type II Weyl semimetal inside the horizon.
 \\
 {\it Left}: Nonequilibrium state in the first moment after formation of the event horizon. 
The positive energy states in the newly formed Fermi surface are fully occupied, while the negative energy states are empty.
The relaxation of this highly nonequilibrium state  is accompanied by Hawking radiation and reconstruction of the vacuum state.
\\
{\it Right}: Final equilibrium state after reconstruction.
The positive energy states are empty, and the negative energy states are occupied. There is no Hawking radiation after this final state is reached. 
}
\label{fig:FS}
\end{figure}
%%%%%%%%%%%%%%%%%%%%%%%%%%%%%%%%%%%%%%%%%%%%%%%%%%%%%%%%%
%%%%%%%%%%%%%%%%%%%%%%%%%%%%%%%%%%%%%%%%%%%%%%%%%%%%%%%%%

In the type II region, where $|v| >1$, the overtilted Dirac cone $E({\bf p}) = p_r v \pm pc$ gives rise to the Fermi surface $E({\bf p})=0$ in Fig. \ref{fig:FS}.  At ${\bf p}=0$ the Fermi surface is terminated by the Weyl point, while  at large ${\bf p}$ the Fermi surface is closed due to the modification of the spectrum in the corresponding ultraviolet (analog of the Planck energy scale), where the nonlinear corrections to the spectrum of fermions become important. In semimetals this scale is typically determined by the interatomic distance.

Such Fermi surfaces also appear in superfluids and superconductors, where they are called the Bogoliubov Fermi 
surfaces.\cite{Volovik1989,Wilczek2003,Agterberg2017,Agterberg2018,Eltsov2019,Sumita2019,Autti2020,Agterberg2021,Timm2021} 

It is important, that  as distinct from fermionic superfluids, in Weyl semimetals the shift velocity in the effective metric is not created by the flow. In fermionic superfluids the Hawking radiation and the corresponding  filling of the Fermi surface lead to the decay of the "black hole".\cite{KopninVolovik1998,JacobsonVolovik1998} In the process of Hawking radiation the flow velocity either decreases until the horizon disappears, or increases due to momentum conservation until the instability threshold is reached and the flow collapses. In both cases the final equilibrium state of the fermionic superfluid does not contain the horizon.
Contrary to that, in Weyl semimetals the fully equilibrium state is finally reached, where the negative energy states in the Fermi surface are fully occupied in  Fig. \ref{fig:FS} ({\it Right}). In this state the horizon is present, while the Hawking radiation is absent. This is the analog of the equilibrium vacuum. In spite of the absence of Hawking radiation, the properties of the event horizon can be studied by scattering of quasiparticles at the horizon, which is also determined by Hawking temperature 

Here we consider the nonequilibrium situation, when the type II region is formed after the rapid variation of the parameters of the system, for example by applying the local pressure. Then at the first moment of creation of the horizon, the state of the system in the overtilted region has inverted population:
 the negative energy states are empty, while the positive energy states are occupied, Fig. \ref{fig:FS} ({\it Left}).
Such "vacuum" state with inverted population corresponds to the negative temperature and is highly nonequilibrium, though in some cases it can represent the ground state. \cite{Volovik2021e}

The initial stage of the process of equilibration is the filling of the negative energy states by the fermions occupying the positive energy states. This process is in particular accompanied by the quantum tunneling from the occupied positivie energy states behind the horizon in the type II region to the empty positive energy states in the type I region. This corresponds to the creation of the particle-hole pairs at the horizon. This is the analog of the Hawking radiation, which is regulated by the Hawking temperature $T_H=v'/2\pi$, where $v'$ is the derivative of the shift velocity at the horizon. Finally the equilibrium "vacuum" state  is formed inside the horizon, in which all  the filled negative energy states inside the Fermi surface become occupied, Fig. \ref{fig:FS} ({\it Right}). In this final vacuum state behind the horizon, there is no Hawking radiation, while the horizon still exists. 

This event horizon is physical, it represents the boundary between two topological phases: the type I and type II semimetals, which have different topology in momentum space.\cite{Volovik2003,Horava2005} The type I semimetal has point nodes in the electronic spectrum (Weyl or Dirac points), which are the Berry phase monopoles.\cite{Volovik1987}
These monopoles are described by the $\pi_3$ topological winding number in the momentum-frequency space. On the other hand, the spectrum of type II semimetal has surfaces of zeroes in momentum space -- the Fermi surfaces, which are protected by the $\pi_1$ topological winding number.\cite{Volovik2003} The boundary between these topological quantum vacua  contains the nodal lines in the electronic spectrum (sometimes this intermediate state with the nodal lines is called the type III Weyl semimetal, see e.g. Ref. \cite{Sims2021}).

\section{Horizon as a quantum phase transition in quantum vacuum}
\label{VacuumStarSec}

Applying this scheme to the black holes (BH), one comes to the following "circle of the life of a BH".\cite{Zubkov2018} At the beginning of its formation,
the BH appears in a non-equilibrium state. This is the conventional state of a BH with the singularity at the origin. This state is quasi-equilibrium, since it is accompanied by Hawking radiation, which leads to relaxation. The interior of the horizon, if it is described by the PG metric, contains the Fermi surfaces.\cite{Huhtala2002,Zubkov2018,Zubkov2018b,Zubkov2019} This determines the further stage of the black hole, when particles start to fill the energy states below the Fermi level.  Since the Fermi surface is not confined, it is in principle impossible to fill all the negative states, which exist up to the Planck energy scale. That is why the vacuum inside  the black hole must be deformed. The filling of negative energy states will be accompanied not only by the Hawking radiation, but also by the back reaction of the gravitational field inside the horizon, which again will be followed by the reconstruction of spectrum. As a result, the black hole interior arrives at the equilibrium state with essentially different properties than the original state.  What state is that?

Here we discuss two possible  equilibrium states of the black hole (if it is still not evaporated by Hawking radiation or by tunneling to the white hole\cite{Rovelli2021,Rovelli2021b,Volovik2021}).
Both configurations contain the de Sitter quantum vacuum in the core of the black hole. The analog of the homogeneous ground state in the type-II semimetal is the state with the homogeneous vacuum energy density, that is the de Sitter state. One of these configurations is the gravastar discussed by Mazur and Mottola,\cite{Mazur2001,Mazur2004,Visser2004} see also Ref. \cite{KlinkhamerVolovik2005} This  gravastar does not have the event horizon. The thin shell of exotic matter exists outside the "event horizon". Let us  call it the type I  gravastar.  In the other possible state, which we call the type II gravsatar, the shell of exotic vacuum is inside the event horizon, see e.g. Refs. \cite{Frolov1990,Dymnikova1992,Dymnikova2002}, the review paper \cite{Dymnikova2020}  and recent Ref. \cite{Maeda2021}. This shell exists between two horizons, outer (event horizon) and inner (Cauchy horizon).

Both firewall objects can serve as an alternative to the conventional black holes, and they can resolve the information loss problem,\cite{Polchinski2013,Hooft2017,Cardoso2019,Hooft2021}  and the puzzles found in the recent observations.\cite{Antoniou2020,Gerosa2021} There is the so-called ”pair-instability” effect, according to which no black holes between $M=52$  and $M=133$ solar mass are expected. However,  the gravitational signal GW190521 shows that there are compact objects within the  forbidden range.

%%%%%%%%%%%%%%%%%%%%%%%%%%%%%%%%%%%%%%%%%%%%%%%%%%%%%%%%%
%%%%%%%%%%%%%%%%%%%%%%%%%%%%%%%%%%%%%%%%%%%%%%%%%%%%%%%%%
\begin{figure}
\centerline{\includegraphics[width=\linewidth]{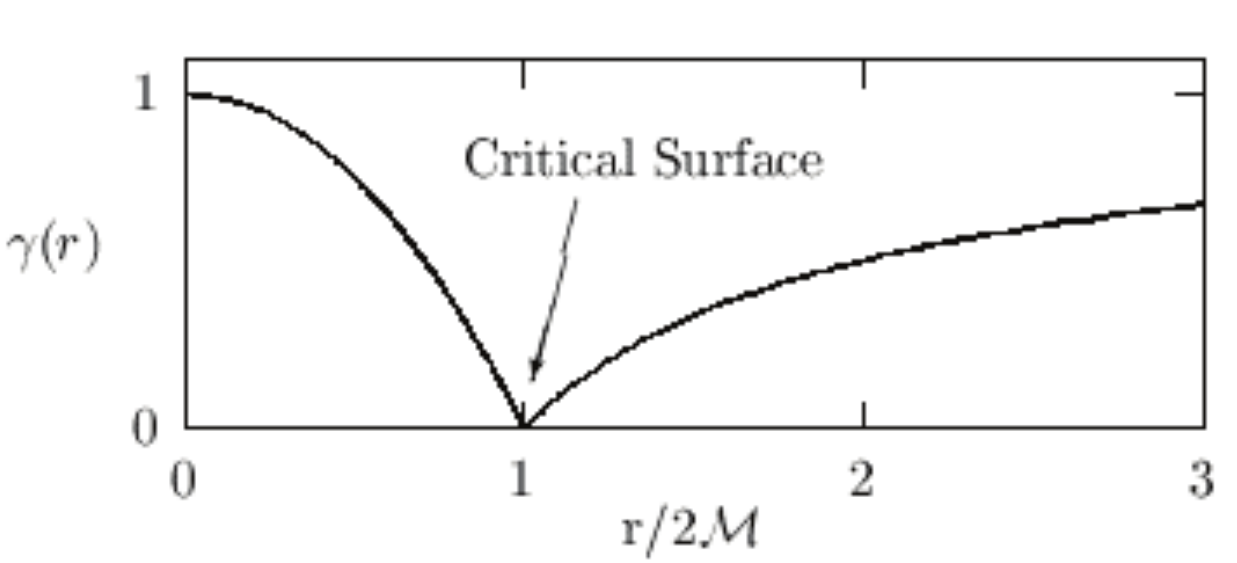}}
\caption{This figure is from Ref.\cite{Chapline2003}. The black hole event horizon is considered therein as a continuous quantum phase transition of the vacuum of space-time.  The function $\gamma(r) = 1 - v^2(r)$.
The "horizon" -- the interface between de Sitter and Schwarzschild  spacetimes --  is at $r=r_h=2M$.
In Weyl semimetals, this is the boundary of the topological quantum phase transtion -- the transition between two different topological vacua.
}
\label{fig:gravastar}
\end{figure}
%%%%%%%%%%%%%%%%%%%%%%%%%%%%%%%%%%%%%%%%%%%%%%%%%%%%%%%%%
%%%%%%%%%%%%%%%%%%%%%%%%%%%%%%%%%%%%%%%%%%%%%%%%%%%%%%%%%

First we consider the intermediate state between the  type I  and type II gravsatars with the singularity just at the horizon, which corresponds to the phase boundary between two vacuum states, see Fig. \ref{fig:gravastar}, which reproduces Fig. 1 in Ref. \cite{Chapline2003}.
The shift velocity in the  PG metric (\ref{PG}) for the singular gravastar has the form:
\begin{eqnarray}
v(r)=- \sqrt{\frac{r_h}{r}} \,\,,\,\, r >r_h \,,
\label{v1}
\\
v(r)=- \frac{r}{r_h} \,\,,\,\, r < r_h \,.
\label{v2}
\end{eqnarray}
Here $r_h=1/(2M)$ is the horizon radius, and $H=-1/r_h$ is the Hubble parameter of the  (collapsing) de Sitter vacuum inside the horizon
(in units $\hbar=G=c=1$).

There is the jump in the velocity gradient  at $r=r_h$, i.e. $v'(r_h-0)=-2 v'(r_h+0)=- 1/r_h $.  The jump $\Delta v'=3/(2r_h)$ at $r=r_h$   gives rise to the singularity in the curvature at the horizon: 
$R \sim \frac{1}{r_h} \delta(r-r_h)$.  The equation of state, which includes the de Sitter vacuum and the region of jump, is anisotropic.\cite{Dymnikova2002,Dymnikova2020} Applying it to the singular metric in Eqs. (\ref{v1}) and (\ref{v2})  one obtains:\cite{Dymnikova2002,Dymnikova2020}  
\begin{eqnarray}
\rho(r) =-p_r(r) =\rho_0 \theta(r_h -r)  \,,
\label{EoS1}
\\
p_\perp(r)= - \rho(r) - \frac{1}{2} r\rho'(r) =- \rho(r)+  \frac{r_h}{2} \rho_0 \delta(r-r_h)\,,
\label{EoS2}
\\
 \rho_0= \frac{3}{8\pi r_h^2}=\frac{3}{8\pi} H^2\,.
\label{EoS3}
\end{eqnarray}
The shift velocity is
\begin{equation}
v^2(r) =2M(r)/r \,,
\label{ShiftVelocity}
\end{equation}
where the mass of the object comes from the energy density $\rho(r)$ of the de Sitter vacuum: 
\begin{equation}
M(r) = 4\pi \int_0^r dx\,x^2 \rho(x)  = \frac{4\pi }{3} \rho_0 r^2 \theta(r_h -r)  + \frac{r_h}{2}   \theta(r-r_h)  \,.
\label{M}
\end{equation}
The singularity in the transverse pressure compensates the  thin shell  curvature singularity in Einstein equation:
\begin{equation}
8\pi p_\perp = -\frac{M''}{r}  
\,.
\label{Einstein1}
\end{equation}
The singular parts of this equation:
\begin{equation}
8\pi p_{\perp \rm{sing}} = -\frac{M''_{\rm{sing}}}{r}   = \frac{3}{2r_h}  \delta(r-r_h)
\,,
\label{Einstein2}
\end{equation}
correspond to the jump $\Delta v'=3/(2r_h)$.

This singularity can be resolved by the construction of the thin layer. This layer can be either inside the event horizon\cite{Dymnikova2020}  or outside the "horizon",\cite{Mazur2001,Mazur2004} which correspond to type II and type I gravastars respectively.
So, the difference between the two types of gravastar is only in the position of the shell, which resolves the singularity: inside or outside the horizon.
If the shell is inside the event horizon  (i.e. between the event and Cauchy horizons), it is made of the vacuum fields. If the shell is outside
the horizon  (i.e. there is actually no event horizon), it is made of the exotic matter fields.

Let us now compare these objects with the  Weyl semimetals.
The type I gravastar is horizonless. In condensed matter  it corresponds to electronic state with the tilted Weyl (or Dirac) cone -- the type I Weyl (or Dirac) semimetal.
The type II gravastar has two horizons. In condensed matter  the state between the two horizons corresponds to electronic state with the overtilted Weyl (or Dirac) cone -- the type II Weyl (or Dirac) semimetal. In condensed matter we have the real boundary between the 
regions with type I and type II Weyl cones. The boundary between these regions is the event horizon, where the corresponding shift vector $|v|=1$ in the proper units. On both sides of the horizon the system is in its ground state -- the analog of the quantum vacuum.
The Hawking radiation is absent, the temperature is zero. 
There are no fermionic quasiparticles -- excitations above the Fermi surface level, i.e. the analog of matter is absent in the full equilibrium
at $T=0$.

So, if the analogy works, it is in favour of the type II gravastar.  In the type II gravastar there is the horizon (even two: the event horizon and the Cauchy horizon). On both sides of the  event horizon at $r=r_h$ there are the vacuum states without matter. The interior of the black hole  contains the thin vacuum shell at $r_h(1-\epsilon) <r < r_h$ with $\epsilon \ll 1$, and the de Sitter vacuum is at $r<r_h(1-\epsilon)$. Within the layer, the shift velocity slightly exceeds the speed of light, $|v|-1 \sim \epsilon$, and the vacuum state represents the occupied Fermi surface with the proper ultraviolet cut-off. The Planck energy cut-off $E_{\rm Planck}$ determines the thickness $1/E_{\rm Planck}$ of the shell. The small excess of the shift velocity above unity within the shell is
$v^2-1 \sim \epsilon \sim 1/(r_h E_{\rm Planck})$;  the singular part of the transverse pressure in the shell in Eq.(\ref{Einstein2}) is
$p_\perp \sim E^3_{\rm Planck}/r_h$; and the surface tension of the shell is $\sigma \propto E^2_{\rm Planck}/r_h$, see also Ref.\cite{Chapline2003}.

\section{Vacuum star in $q$-theory}
\label{VacuumQSec}

Due to the Planck length scale, the consideration in terms of the Fermi surface does not look promising.
Such scale suggests that the consideration of the shell at $r_h(1-\epsilon) <r < r_h$ should be done in terms of the vacuum fields. 
One may consider the inhomogeneous vacuum state in  the type II gravastar using the variables ($q$-fields) describing the phenomenology of the quantum vacuum,\cite{KlinkhamerVolovik2008a,KlinkhamerVolovik2008b} such as the 4-form field\cite{Duff1980,Aurilia1980,Hawking1984} or the tetrad determinant, which is also the 4-form,\cite{Volovik2020c} see also Ref. \cite{Guendelman2021}.
This approach allows us to describe the deep quantum vacuum without knowing the microscopic theory at the Planck scale.
The quantum vacuum in this description is the self-sustained system, which adjusts itself to the external conditions. The black hole exterior with fixed mass $M$ provides the external condition for the quantum vacuum inside the horizon.

The action for the vacuum field $q$ interacting with gravity has the form similar to that for the scalar field:\cite{KlinkhamerVolovik2017b,KlinkhamerVolovik2017}
 \begin{eqnarray}
S=\int d^4x \sqrt{-g}\left(\frac{R}{16\pi G} +\epsilon(q)
+\frac{1}{2}\, g^{\alpha\beta}\, \nabla_\alpha q \nabla_\beta q \right).
\label{eq:action}
\end{eqnarray}
However, the main difference of the 4-form field from the scalar field is that in the Einstein equation the energy density of the $q$-field, $\epsilon(q)$, is shifted by the chemical potential term , $\epsilon(q) \rightarrow \rho(q)= \epsilon(q) -\mu q$:\cite{KlinkhamerVolovik2017b,KlinkhamerVolovik2017}
\begin{eqnarray}
\nonumber
R_{\alpha\beta} - \frac12\, g_{\alpha\beta}\,R =   
\\
=g_{\alpha\beta} \left(\rho(q)+ \frac{1}{2} \nabla_\alpha  q \nabla^\alpha q  \right)
  + \nabla_\alpha  q \nabla_\beta  q \,,
  \label{eq:Einstein-eq}
  \\
\rho(q)=\epsilon(q) - \mu q  \,.
  \label{eq:rho}
\end{eqnarray}
This allows to solve the main cosmological constant problem: in the full equilibrium vacuum the huge Planck scale energy density 
$\epsilon(q_0) \sim E_{\rm Planck}^4$ is compensated by the term  $\mu q_0$ without any fine tuning: $ \rho(q_0)= \epsilon(q_0) -\mu q_0=0$. Such compensation takes place not only in the relativistic vacuum, but is well known in condensed matter, where the role of $q$ is played by the particle density $n$: at zero external pressure and at zero temperature one has
$\epsilon(n_0)-\mu n_0=0$. This follows from the Gibbs-Duhem thermodynamic relation, which is universal and thus works in the full equilibrium irrespective of whether the system relativistic or not.

The action (\ref{eq:action}) suggests the following distribution of the vacuum field in the three regions:
outside the event horizon, $r>r_h$; in the de Sitter core of the black hole inside the Cauchy horizon, $r<r_h(1-\epsilon)$; and in the vacuum shell  between the two horizons, $r_h>r>r_h(1-\epsilon)$.

Outside the event horizon,  $r>r_h$,  the vacuum variable $q$ has the equilibrium value $q_0 \sim E_{\rm Planck}$, which corresponds to zero value of the cosmological constant, $\rho(q_0)=0$. This fixes the value of the chemical potential $\mu = d\epsilon/dq|_{q=q_0}$, which is constant both outside and inside the event horizon.

Inside the event horizon,  at $r<r_h$,  the deviation of $q$ from its equilibrium value is small, $|q-q_0| \ll q_0 \sim E_{\rm Planck}$, and thus $\rho(q)$ can be expanded in terms of deviations:
\begin{eqnarray}
\rho(q) =  \epsilon(q) - \mu q  
\label{deviations1}
\\
=\rho(q_0) +  \left(d\epsilon/dq|_{q_0} -\mu \right)(q-q_0) + \frac{1}{2} \frac{d^2\epsilon}{dq^2} (q-q_0)^2 
\label{deviations2}
\\
=  \frac{1}{2} \frac{d^2\epsilon}{dq^2} (q-q_0)^2  \sim q_0^2 (q-q_0)^2
\,.
\label{deviations3}
\end{eqnarray}
Here we used $\rho(q_0) =  \epsilon(q_0) - \mu q_0=0$ and $d\epsilon/dq|_{q=q_0}=\mu$.

In the de Sitter region, $r<r_h(1-\epsilon)$, the vacuum variable $q$ is constant with $(q-q_0)^2 = 1/r_h^2$. This corresponds to the vacuum energy  density $\rho \sim E_{\rm Planck}^2/r_h^2$, which forms the de Sitter vacuum inside the black hole with $H^2=1/r_h^2$.

Let us now consider the vacuum shell,  $r_h>r>r_h(1-\epsilon)$. The shell width is determined by the characteristic length of the $q$-field. It can be obtained by comparing of the energy density in Eq.(\ref{deviations3}) with the gradient term $(\nabla q)^2$ in Eq.(\ref{eq:Einstein-eq}). The characteristic length is on the order of Planck length  $l_{\rm Planck}=1/E_{\rm Planck}$, and thus $\epsilon=l_{\rm Planck}/r_h \ll 1$. 
Since the $\delta$-function in Eq.(\ref{Einstein2}) has the  Planck scale width, the vacuum energy density within the shell is on the order of $\rho =E^3_{\rm Planck}/r_h$. This gives the estimation of the deviation of the vacuum field from its equilibrium value within the shell:  $(q-q_0)^2 \sim  E_{\rm Planck}/r_h$. This deviation is still small compared with the Planck energy scale, $|q-q_0| \ll q_0 \sim E_{\rm Planck}$. The surface tension of the vacuum shell is $\sigma \propto E^2_{\rm Planck}/r_h$, which is in agreement with Ref.\cite{Chapline2003}.

\section{Discussion}

In this construction of the vacuum star, which we call here  the type II gravastar, there is no conventional or thermal matter.  The interior of the black hole inside the event horizon is fully expressed in terms of the vacuum fields. The Hawking radiation is absent in this final equilibrium state just for the same reason as in the Weyl semimetals,\cite{Volovik2016} after all the negative electronic energy states in the Fermi surface emerging inside the horizon are occupied. In the final equilibrium state there is no dissipation and thus no Hawking radiation. The event horizon (and also the Cauchy horizon) represents the physical surface, which separates the vacua with different behavior of the vacuum variable. This is also in agreement with the physical horizon in Weyl semimetal, which separates  vacua with different momentum-space topology.
 
Such vacuum construction is distinct from the type I gravastar in the Mazur and Mottola model,\cite{Mazur2001,Mazur2004}  where the layer of positive pressure fluid is just outside the 'event horizon'. In  the type II gravastar the layer does not represent the matter field, and thus its equation of state (EoS) may have any form including EoS $p=\rho$ of stiff fluid suggested by Zeldovich\cite{Zeldovich1962} and used by  Mazur and Mottola.\cite{Mazur2001,Mazur2004}   Note that in the $q$-theory, the space-time variation of the vacuum fields  mimics the cold dark matter in expanding Universe.\cite{KlinkhamerVolovik2017} 

It would be interesting to apply this approach to the fate of the rotating (Kerr) black hole.\cite{Mottola2021} It is possible that the black hole may reach the quasi-equilibrium state under rotation. In this state of the Kerr black hole, the Hawking radiation is absent, while the rotational friction will be 
due to the Zeldovich-Starobinsky effect -- the amplification and spontaneous emission of 
waves by any body rotating in the quantum vacuum, including the Kerr black hole.\cite{Zeldovich1971,Zeldovich1971b,Starobinskii1973,Takeuchi2008}  

In conclusion, we considered the possible scenario of the formation of the equilibrium final state of the black hole. This scenario is inspired by the consideration of the black hole analog in Weyl semimetals, where the analog of the event horizon separates two topologically different types of the Weyl materials: type I and type II. The relaxation of the initial state of the black hole analog is accompanied by the Hawking radiation and by the reconstruction of vacuum state inside the horizon. In the final equilibrium state, the event horizon separates two different vacuum states, and there is no more Hawking radiation. Such final state in Weyl semimetals is analogous to the dark energy stars discussed earlier. As distinct from the gravastar, in such black hole there is the real event horizon. The interior of the event  horizon contains the de Sitter vacuum and the thin singular shell, both made of the vacuum fields.
We call this type of the vacuum stars as the type II gravastar. We suggest to use the $q$-theory of the quantum vacuum to study the internal structure of such compact objects.

  {\bf Acknowledgements}.  I thank M. Zubkov for discussions. This work has been supported by the European Research Council (ERC) under the European Union's Horizon 2020 research and innovation programme (Grant Agreement No. 694248).

\end{document}